\documentclass[aps,prl,showpacs,twocolumn,groupedaddress]{revtex4}

\usepackage{graphicx}	

\newcommand{\calD}{{\cal D}}

\begin{document}


\title{Dynamics of one-dimensional Bose liquids: \\
Andreev-like reflection at Y-junctions and absence of the Aharonov-Bohm effect}


\author{Akiyuki Tokuno$^{1,2}$, Masaki Oshikawa$^1$, Eugene Demler$^3$ }
\affiliation{ $^1$Institute for Solid State Physics, University of Tokyo,
Kashiwa 227-8581 Japan \\
$^2$ Department of Physics, Tokyo Institute of Technology,
Oh-okayama, Meguro-ku, Tokyo 152-8551 Japan \\
$^3$ Department of Physics, Harvard University, MA 02138}


\date{\today}

\begin{abstract}
We study one dimensional Bose liquids of interacting ultracold atoms
in the Y-shaped potential when each branch is filled with
atoms. We find that the excitation packet incident on a single
Y-junction should experience a negative density reflection analogous
to the Andreev reflection at normal-superconductor interfaces,
although the present system does not contain fermions.  In a ring
interferometer type configuration, we find that the transport is
completely insensitive to the (effective) flux contained in the ring,
in contrast to the Aharonov-Bohm effect of a single particle in the
same geometry.
\end{abstract}

\pacs{03.75.Kk}

\maketitle

Recently,  guiding of atoms in
low-dimensionally magnetic traps
has been actively studied~\cite{OFSGZ}.
Such systems provide an ideal opportunity to study coherent
quantum dynamics of interacting many-body systems.
The theory of one-dimensionally trapped atoms in equilibrium
~\cite{Sen,Cazalilla1,Cazalilla2,BGOL},
as well as questions related to coherent dynamics and nonequilibrium 
phenomena~\cite{JS1}, has been vigorously studied.
In this Letter we analyze a simple yet nontrivial example of
the real-time dynamics of Bose-Einstein condensates (BEC) in the
Y-shaped potentials (beam-splitters)~\cite{JS2},
and in the related ``ring interferometer'' type geometry~\cite{ATFAHS}.

\begin{figure}[tbp]
 \begin{center}
  \includegraphics[scale=0.5]{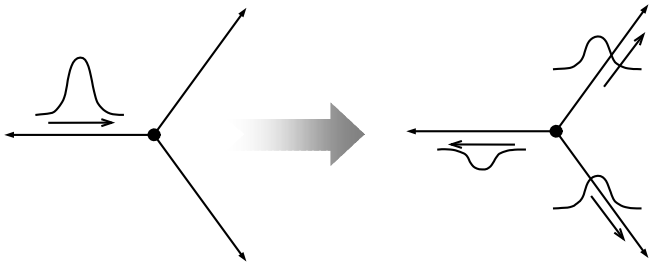}
  \caption{Reflection and the transmission of the injected
  current. The left panel describes the initial situation,
  and  the right panel shows the situation after the
  reflection and the transmission. \label{fig2}}
 \end{center}
\end{figure}

In contrast to previous works where the
dynamics of wave packets in otherwise empty traps is studied,
here we consider the situation where each one-dimensional branch of the
Y-shaped potential is {\em filled with interacting bosonic atoms}, i.e.
each branch contains one-dimensional Bose liquid.
The Tomonaga-Luttinger (TL) liquid theory~\cite{Haldane,Cazalilla1}
is a powerful method to study this problem.
The whole system in the Y-shaped potential
may be regarded as three TL liquids
connected at a junction, as in Fig.~\ref{fig2}. 
While there are parallels with the corresponding electron
problem~\cite{NFLL,COA}, as we will show in the following, 
a quite different generic behavior is expected in the Bose liquid.
In the present case, instead of the conductance,
the real-time dynamics could be directly observed
in experiments.
Furthermore, as an application of the above analysis of
the single Y-junction problem,
we also consider the transport in a ring type interferometer geometry.

We have two main results in this Letter.
The first is the prediction of the negative density
reflection at a junction (Fig.~\ref{fig2}). In the discussion of Y
junctions for systems of interacting electrons, such negative reflection
has been interpreted as the Andreev reflection,
which occurs in metal-superconductor interfaces due to
Cooper pair formation.
On the other hand, the negative reflection at the BEC Y-junction
can be understood rather simply using
an analogy with electromagnetic transmission line three-way junction.
For an ideal three-way junction, there should be no voltage difference
between the connected ends of the three lines, and there is conservation
of the total charge in a pulse. These two conditions imply that when the
amplitude of an incident signal is one, signals transmitted into the
other two lines will have amplitudes of $2/3$ and the reflected signal
will have the amplitude of $-1/3$. As we discuss below the same
phenomenon should appear in a BEC Y-junction.



\begin{figure}[tbp]
 \includegraphics[scale=0.7]{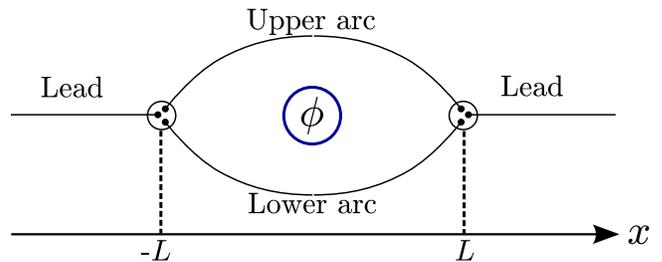}
 \caption{The schematic figure of the ring interferometer made
of two Y junctions, with left and right leads, where $\phi$ is an
 effective ``magnetic'' flux enclosed by the ring. \label{fig3}}
\end{figure}

The other main result is the absence of the Aharonov-Bohm (AB)
effect in the ring type interferometer, which consists of two
Y-shaped junctions as in Fig.~\ref{fig3}.
As a typical example of the AB effect, when the ring is empty,
a single particle injected from one lead to the
ring does not transmit to the other lead when the (effective)
``magnetic flux'' inside the ring
is the half unit flux quantum. However, we find
that the transport between the leads in our system
is completely insensitive to the magnetic flux, in the low energy limit.
These two findings are manifestation of the collective nature
of the Bose liquid, rather than single-particle physics.

%


We start from the following one-dimensional Hamiltonian~\cite{Olshanii}
with the low-energy scattering characterized by the low-energy s-wave
scattering:
\begin{equation}
 {\cal H}_0
 =\sum_{j=1}^{3}\int\!\!\! dx
  \left[
   \frac{1}{2m}\partial_{x}\psi_j^{\dagger}(x)\partial_{x}\psi_j(x)
   +\frac{U}{2}\rho_j^2(x)
  \right], \label{HTL}
\end{equation}
where we take an unit of $\hbar=1$. 
$\psi_j(x)$, $\rho_j(x) \equiv \psi_j^{\dagger}(x) \psi_j(x)$
are, respectively, the annihilation and the density operator
of the boson with the mass $m$, on the $j$-th branch. ($j=1,2,3$.)
The effective interaction strength $U>0$ is determined by the s-wave
scattering length. 
We take $x>0$, so the physics at the junction is represented by the
boundary condition (b.c.) at $x=0$.
In case of the junction problem, the b.c. is a non-trivial
problem and plays a crucial role. First of all, we
have to identify what kind of b.c. describes the
physics in the low-energy limit.

Any b.c. should satisfy the total
current conservation
\begin{equation}
  J_1(t,x=0)+J_2(t,x=0)+J_3(t,x=0)=0
  \label{CurrCons}
\end{equation}
where $J_j(x)$
is the current operator of the Bose liquid in the $j$-th
branch. The simplest possible b.c. corresponds to
an infinitely strong barrier at the junction, which
prohibits any particle to be transmitted through the junction.
It would give $J_j(t,x=0) = 0$ for $j=1,2,3$.
Naturally, this satisfies the current conservation~(\ref{CurrCons}).

In practice, there would be a finite tunneling amplitude $\Gamma$
between the different branches.
The effect of the tunneling can be included by adding
the perturbation
\begin{equation}
{\cal H}_B=-\sum_{j=1}^{3} \left[\Gamma
	    \psi_j^{\dagger}(x=0)\psi_{j-1}(x=0)+\mbox{h.c.} \right],
\label{Hboundary}
\end{equation}
at the boundary.
We are interested in the low-energy behavior
of the junction in the presence of the tunneling.

In order to study the low-energy physics,
it is convenient to use the ``bosonization'' technique~\cite{Haldane}.
In terms of the TL boson field $\theta_j(x)$
and its dual $\varphi_j(x)$,
the Bose atom annihilation operator and density operator
on the $j$-th branch are, respectively, represented as
$\psi_j^{\dagger}(x)\sim\left[\rho^{(0)}+\partial_x\theta_j(x)/\pi\right]^{1/2}e^{-i\varphi_j(x)}$,
$\rho_j(x)\sim\rho^{(0)}+\partial_x\theta_j(x)/\pi$,
where we have retained only the leading terms.
$\rho^{(0)}$ represents the expectation value of the atom
density in the ground state.
The low-energy effective Hamiltonian for equation~(\ref{HTL}) is given by
\begin{equation}
 {\cal H}_{0}=\sum_{j=1}^3 \frac{v}{2\pi}\int_0^L\!\!\! dx
  \left[g(\partial_x\varphi_j)^2
  +g^{-1}(\partial_x\theta_j)^2\right].
  \label{HTL2} 
\end{equation}
The fields $\theta$, $\varphi$ satisfy
the commutation relation $[\theta(x), \varphi(y)]=i\pi H(x-y)$, where
$H(x-y)$ is the Heaviside step function. $v$ is the velocity of the
collective modes, and $g$ is the so-called TL parameter,
in which
the interaction
is essentially contained.

The TL parameter, for the interacting fermions i.e. electrons, 
usually satisfies $g<1$ due to the Coulomb repulsion.
However, $g>1$ occurs rather naturally in a Bose liquid
with a short-range repulsive interaction.
For example, for the Bose liquid with a $\delta$-function
interaction~\cite{LL},
the TL parameter $g$ can be determined from the
Bethe Ansatz exact solution.
It turns out that $g$ is always larger than $1$ for this model,
interpolating between $g=\infty$ in the non-interacting limit
and $g=1$ in the strongly interacting (or dilute)
``Tonks gas'' limit.~\cite{Cazalilla1,Cazalilla2,BGOL}
While $g$ can also be less than $1$
with other types of the interaction~\cite{AHNS},
$g>1$ would be rather generic for the interacting Bose liquid.

In terms of the bosonization, the ``disconnected'' limit
$\psi_j(0)=0$ corresponds to the Dirichlet boundary
condition $\partial_t\theta_j(0)=0$ for $j=1,2,3$.
It is equivalent to the Neumann b.c. on
$\varphi_j$. The tunneling perturbation (\ref{Hboundary})
is bosonized as ${\cal H}_{B} \sim \Gamma
e^{-i(\varphi_j(0)-\varphi_{j-1}(0))}$,
where only the most relevant term is retained.
The scaling dimension of this perturbation
is given by $g^{-1}$, and it is relevant when $g>1$.
Namely, the effective tunneling amplitude $\Gamma$ grows as
the energy scale such as the temperature is lowered.
It is expected that $\Gamma$ eventually grows to infinity
in the low-energy limit, when $g>1$.~\cite{Saleur}

In the strong coupling limit, the $\varphi$ fields should be pinned as
$\varphi_1(t,0)=\varphi_2(t,0)=\varphi_3(t,0)$,
because the tunneling term acts as an infinitely strong potential
on $\varphi$ at the junction.
This also implies
$\partial_t\varphi_1=\partial_t\varphi_2=\partial_t\varphi_3$.
Translating it into the b.c.s on $\theta_j$, we find
\begin{eqnarray}
 && \sum_{j=1}^3 \partial_t \theta_j (t,x=0)=0, \label{CurrConv2} \\
 && \partial_x \theta_i(t,x=0)=\partial_x\theta_j(t,x=0) \ (i,j=1,2,3), \label{BC2}
\end{eqnarray}
where the first line comes from the current conservation~(\ref{CurrCons}).
The second line indicates that the density is continuous at the
junction, reflecting that the junction is at the strong coupling limit.
Using the boundary conformal field theory techniques,
we can see that all the possible perturbations to
the above b.c. are irrelevant if
$g> 3/4$~\cite{NFLL,COA} in the case of Bose liquid.
This guarantees the stability of the ``strong coupling'' fixed point,
supporting that $\Gamma$ grows to infinity in the low-energy limit
when $g > 1$.

 

Now we discuss the dynamics of the system with the
stable, strongly coupled b.c..
Let us suppose that the density 
is altered
locally from the equilibrium.
The variation of the density propagates in the system,
and eventually would be transmitted to other branches.

We start from the
Heisenberg equation of motion for the TL boson field $\theta_j$
and dual field $\varphi_j$ ($j=1,2,3$):
$\partial_t \theta_j(t,x)=-vg\partial_x \varphi_j(t,x)$,
$\partial_t \varphi_j (t,x)=-\frac{v}{g}\partial_x \theta_j(t,x)$.
Let us focus on the expectation values of the density and
current. 
Instead of {\em operators} $\theta_j$ and $\varphi_j$,
we then just need to determine the expectation values
$\langle\theta_j\rangle$ and $\langle \varphi_j \rangle$,
which obey the simple one-dimensional {\em classical wave equation}.
The initial conditions are given as
the expectation values of current and density at $t=0$.

For simplicity, we consider
the situation where a left-moving packet is injected from one side (branch 1).
Since the left movers go towards the boundary ($x=0$),
and the right movers go away from the boundary,
the initial conditions may be given as
$\langle\rho_j^L(t=0,x)\rangle=\rho^{(0)}+\calD_0(x)\delta_{1,j}$,
$\langle\rho_j^R(t=0,x)\rangle=\rho^{(0)}$,
where $\calD_0\equiv\partial_x\langle \theta^L_1(t=0,x)\rangle/\pi$ is
the scalar which gives the initial condition on the expectation value of
the densities.
Solving the classical wave equation with the initial conditions,
we obtain:
\begin{eqnarray}
 \langle \rho_1(t,x) \rangle
  &=&\rho^{(0)}+ \calD_0(\eta^+)-\frac{1}{3}\calD_0(\eta^-)H(\eta^-),
\label{C1}\\
 \langle \rho_{2,3}(t,x) \rangle
  &=&\rho^{(0)}+\frac{2}{3}\calD_0(\eta^-)H(\eta^-),
\label{C2}
\end{eqnarray}
where $\eta^{\pm} \equiv vt \pm x$.
The second term in the rhs of the equation (\ref{C1}) represents the incident
packet density, and the third term implies the (negative)
reflection at the boundary.
Namely, if a bump in the density moves through the branch $1$
to the junction, a dip in the density is reflected back. (see Fig.\ref{fig2})

Eq.~(\ref{C2}) shows the
transmission density from the branch $1$.
Defining the transmission tensor as the ratio of
the incident current and transmission currents, the
transmission tensor from the branch $1$ to the branch $2$ (branch
$3$), $T_{2,1}$ ($T_{3,1}$), are obtained as $T_{2,1}=T_{3,1}=\frac{2}{3}$.
This means that the sum of the densities transmitted to branches
$2$ and $3$ is greater than the incident density from the branch $1$.
The current conservation law is satisfied with the
``negative reflection'' current in the branch $1$, which
appears as the second term in equation (\ref{C1}).
The time evolution of the density (\ref{C1})--(\ref{C2}),
including the remarkable negative reflection,
could be observed in experiments, with present techniques
on atomic BEC.



A similar behavior was found in the Y-junction of one-dimensional
electron systems, in terms of the conductance.
It was argued as a consequence of Andreev reflection similar to
that in a superconductor-normal interface.
What is surprising here is that the Andreev-like negative reflection
is also found for the boson systems.
The system of bosons is usually not related
to Cooper pair formation, which is the underlying mechanism
of the Andreev reflection.
These results may be also derived from a linearized mean-field theory
which is applicable to the weakly interacting regime. While the
mean-field approximation breaks down in the strongly interacting regime,
our derivation based on the TL liquid theory are valid even in the
presence of strong fluctuation. Thus our results represent universal
feature of general repulsively interacting boson systems.



Let us extend our analysis to a ``ring interferometer'' type configuration,
which consists of two Y-junctions. (see Fig.\ref{fig3})
We take the TL fields on the upper and lower arc in the ring to
$\theta_{u,l}$, respectively. Then the symmetric combination
$\theta_s=\theta_u+\theta_l$ is coupled with the right
and left leads,
while anti-symmetric combination $\theta_a=\theta_u-\theta_l$ is
decoupled. Here, the propagation of the atoms in the ring
are symmetric so that anti-symmetric combination does not affect
the propagation. As a
result, we can obtain the inhomogeneous TL liquid ~\cite{TLconductance} as
the effective model of the double Y junction systems, where the TL
parameter on the left and right lead is $g$, and that of the symmetric
combination $\theta_{s}$ is $2g$. Following Ref.~\cite{SS2},
we can investigate the transmission and the reflection at the junctions.
The solution of the equation of motion naturally gives the contribution
of multiple reflections. 
Interestingly, the effect of normal and
Andreev-type reflections cancel out, so that the total transmission
approaches $1$ as the elapsed time approaches infinity.
Experimentally, real-time observations of the density
could resolve each reflection.

Now let us consider the effect of a ``magnetic flux'' $\phi$ inside the ring.
The present geometry is a typical setting to observe the AB effect.
If we consider the transport of a single charged
particle in the same geometry with an empty ring, 
the transmission probability from the left lead to the right 
vanishes when $\phi$ is the half unit flux quantum.
This is because of the cancellation between
the probability amplitudes coming from the two paths.
Although a magnetic flux actually does not induce the AB effect for
neutral atoms,
several possible approaches for introducing gauge fields and AB type
phase factors for neutral atoms have already been discussed in the
literature~\cite{ZFWZD}. A common ingredient of many methods is that of
the Berry's phase arising from the orbital motion of
atoms~\cite{JORK}.


We find that, however, in our problem of {\em filled}
one-dimensional BEC,
the transport between the leads is completely insensitive to
the effective magnetic flux $\phi$, in the low-energy limit.
Namely, the AB effect is absent.
This rather surprising conclusion follows
quite naturally from the present formulation, as follows.

With a gauge transformation,
the effect of the flux can be described by the boundary
conditions $\varphi_u(-L)=\varphi_l(-L)$ at the left junction
and $\varphi_u(L)=\varphi_l(L)+\phi$ at the right one.
They determine the boundary values of the antisymmetric
combination as $\varphi_a(-L)=0$ and $\varphi_a(L)=\phi$.
The current on the ring is just a superposition of the symmetric flow between
the two leads described by $\varphi_s$ and the persistent current
described by $\varphi_a$. Only the latter depends on the flux $\phi$.
As it is only the symmetric combination $\varphi_s$ that
couples with the leads, the flux $\phi$ has no
effect on the transmission between the leads.

If we consider the system with
fermions instead of bosons in the same geometry,
the AB effect must be present,
at least when the fermions are non-interacting.
As the non-interacting fermions in one dimension
corresponds to the TL liquid with $g=1$, our result might appear
in contradiction.
However, there is actually no contradiction, because our analysis
applies only to the ``strongly coupled'' b.c. (\ref{BC2}).
In the system of non-interacting electrons, the junction is described
by a different b.c..
The ``strongly coupled'' b.c. is expected to be stable
even in a fermionic system, when the interaction is sufficiently
attractive ($g>9$)~\cite{COA}.
For such a system, the AB effect should be absent as well. 

Finally, we discuss the limitations of and possible corrections to
the present results.
First, the TL liquid description is only valid in the low-energy limit.
This requires $T, 2\pi v/L \ll \mu$, where $L$ is the system size
(length of the branches), and the chemical
potential $\mu = U \rho^{(0)}$ gives
the ultraviolet cutoff.~\cite{Cazalilla1}
When a packet of extra particles is injected, the variation
$\Delta \rho$ of the particle density
corresponds to a variation of the local chemical potential
$\Delta \mu = U \Delta \rho$.
Thus $\Delta \rho \ll \rho^{(0)}$ is required for the TL liquid
description to apply.
For a larger amplitude, the single-particle physics is expected
to show up, instead of the collective nature characterized
by the TL liquid.

The ``strongly coupled'' b.c.~(\ref{BC2}), which
we assumed in the present Letter, is exact
only in the low-energy limit, even when the bulk is completely
described by the TL liquid.
There are corrections due to finite energy scales present in
a realistic system.
The leading correction
to the results based on
the b.c.~(\ref{BC2}) can be related to,
at a sufficiently low energy scale,
the leading irrelevant operator
with scaling dimension $4g/3$.
It would give, for example, a positive reflection proportional to
$ \epsilon^{8g/3 - 2} $ on top of the universal negative
reflection $-1/3$.
Here $\epsilon$ denotes the largest among the possible
infrared cutoff scales, such as $T$, $2\pi v/L$, and $U \Delta \rho$.
When the interaction is weak, it is difficult to satisfy
the conditions for the TL liquid theory
to be applicable in the bulk.
Interestingly, however, the corrections at the junction vanishes
quickly because $g$ is large in such a case.


\bigskip

The authors would like to thank I. Affleck, C. Chamon and C.-Y. Hou for 
useful discussion on related problems, and S. Furukawa
for a valuable comment on the absence of the AB effect.
A. T. was supported by JSPS. 
This work was supported in part by 21st Century COE program
at Tokyo Institute of Technology `Nanometer-Scale Quantum Physics'
from MEXT of Japan, the NSF grant No. DMR-0132874,
AFOSR, and the Harvard-MIT CUA.

\end{document}